\documentclass[runningheads,a4paper]{llncs}

\usepackage{amssymb}
\usepackage{graphicx}

\usepackage{changepage}
\usepackage{subfig}

\usepackage{url}
\urldef{\mailsa}\path|{alfred.hofmann, ursula.barth, ingrid.haas, frank.holzwarth,|
\urldef{\mailsb}\path|anna.kramer, leonie.kunz, christine.reiss, nicole.sator,|
\urldef{\mailsc}\path|erika.siebert-cole, peter.strasser, lncs}@springer.com|
\newcommand{\keywords}[1]{\par\addvspace\baselineskip
\noindent\keywordname\enspace\ignorespaces#1}
\newcommand*{\affaddr}[1]{#1} 
\newcommand*{\affmark}[1][*]{\textsuperscript{#1}}

\begin{document}

\mainmatter  

\title{Extraction of Coronary Vessels in \\Fluoroscopic X-Ray Sequences Using\\Vessel Correspondence Optimization}

\titlerunning{Coronary Artery Extraction Using Vessel Correspondence Optimization}

%
%
\author{
Seung Yeon Shin\affmark[1\dag]
\and Soochahn Lee\affmark[2\ddag] \and Kyoung Jin Noh\affmark[2]\and \\Il Dong Yun\affmark[3] \and Kyoung Mu Lee\affmark[1] \thanks{\footnotesize {This work was supported by the Institute for Information \& communications Technology Promotion(IITP) Grant (No.R0101-16-0171) and by the National Research Foundation(NRF) Grant (2015R1A5A7036384), both funded by the Korean Government(MSIP).}}
}

%
\authorrunning{S.Y. Shin et al.}



\institute{\affaddr{\affmark[1]Dept. ECE, ASRI, Seoul Nat'l Univ.},
\affaddr{\affmark[2]Dept. Electronic Eng., Soonchunhyang Univ.},\\
\affaddr{\affmark[3]Div. Comp. \& Elec. Sys. Eng., Hankuk Univ. of Foreign Studies}\\
\email{\affmark[\dag]syshin@snu.ac.kr, \affmark[\ddag]sclsch@sch.ac.kr}
}

%
%

\toctitle{Coronary Artery Extraction Using Vessel Correspondence Optimization}
\tocauthor{S.Y. Shin et al.}
\maketitle

\begin{abstract}
We present a method to extract coronary vessels from fluoroscopic x-ray sequences. Given the vessel structure for the source frame, vessel correspondence candidates in the subsequent frame are generated by a novel hierarchical search scheme to overcome the aperture problem. Optimal correspondences are determined within a Markov random field optimization framework. Post-processing is performed to extract vessel branches newly visible due to the inflow of contrast agent. Quantitative and qualitative evaluation conducted on a dataset of 18 sequences demonstrate  the effectiveness of the proposed method.
\end{abstract}

\section{Introduction}

\begin{figure}[t]
\centering
\subfloat{\includegraphics[width = 0.24\linewidth]{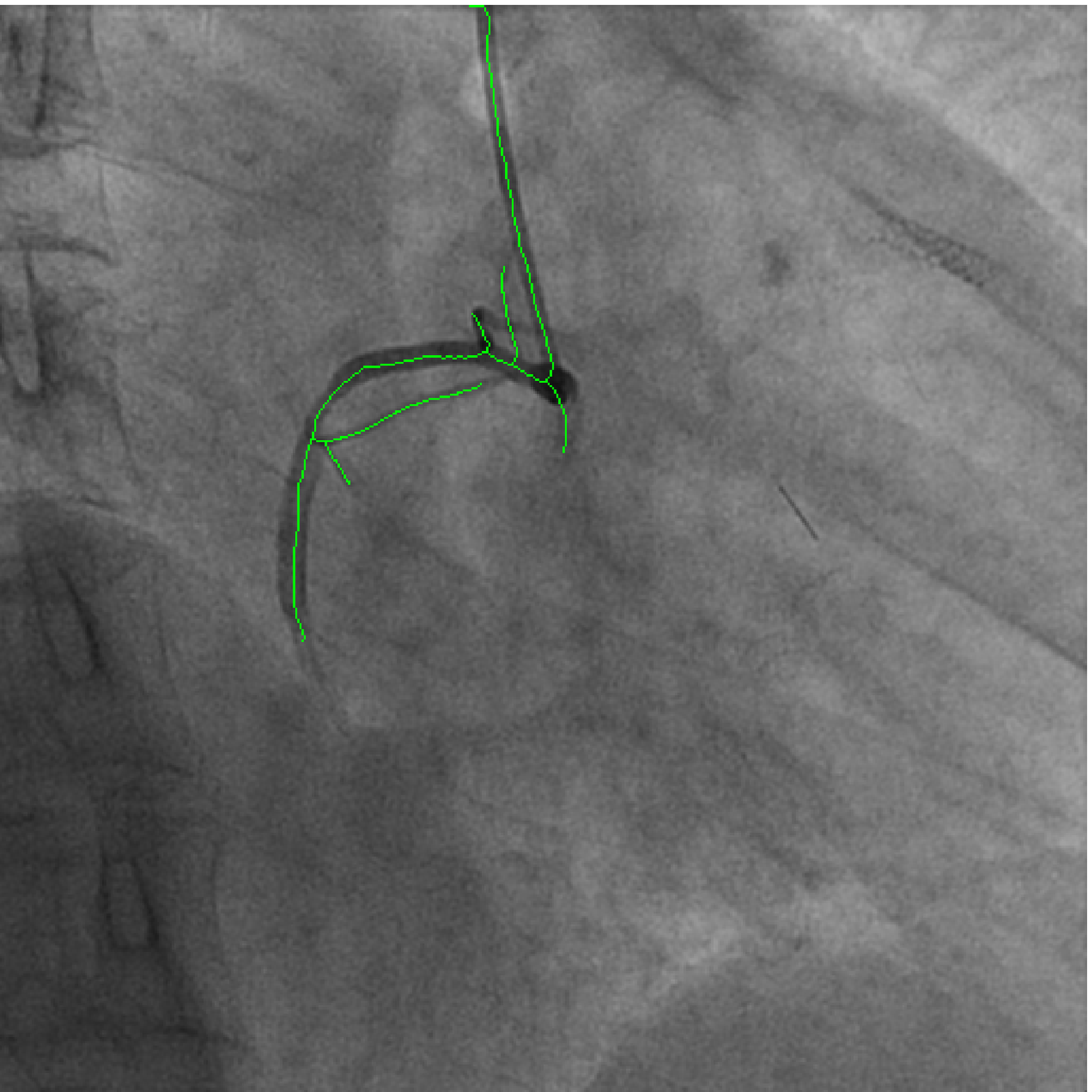}}
\subfloat{\includegraphics[width = 0.24\linewidth]{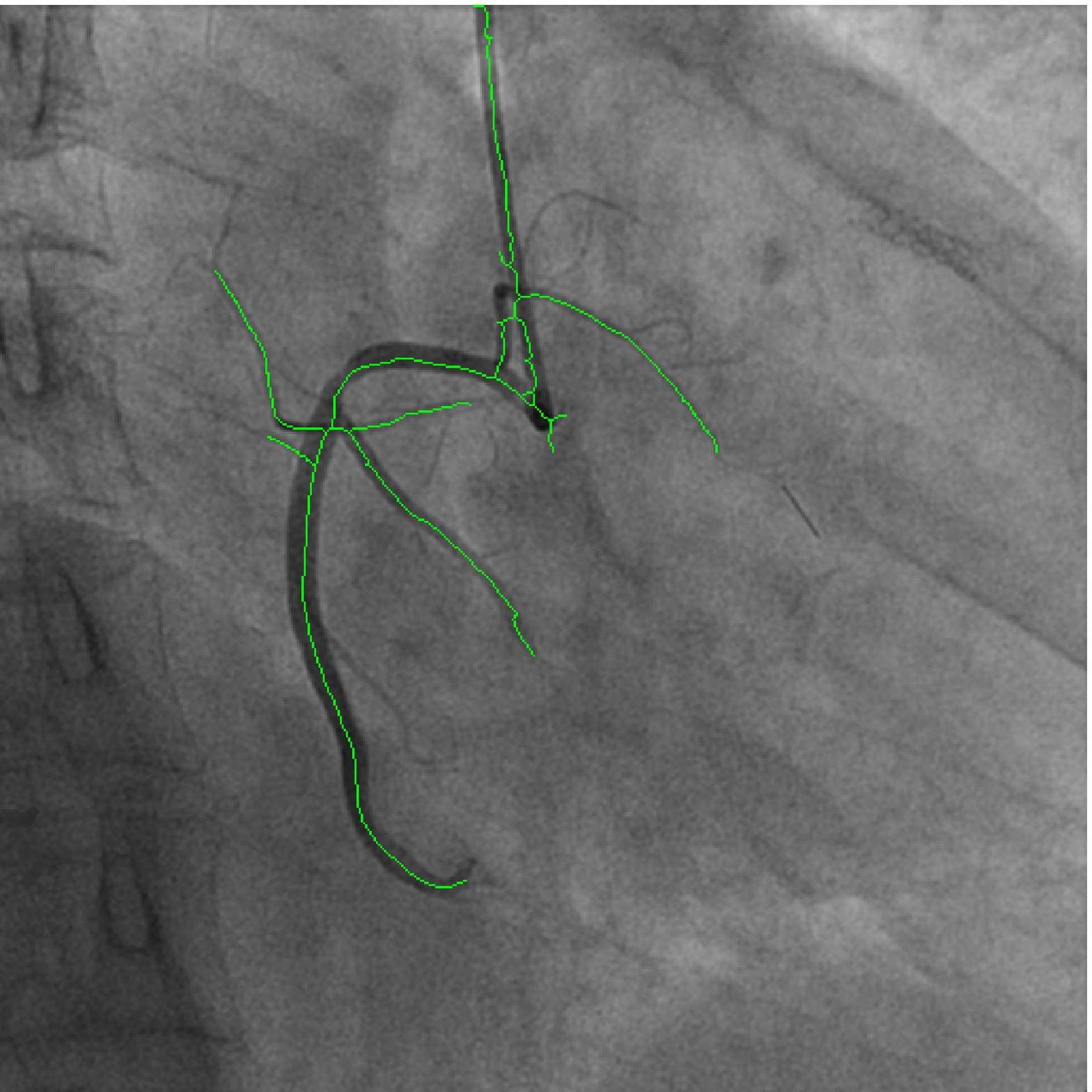}}
\subfloat{\includegraphics[width = 0.24\linewidth]{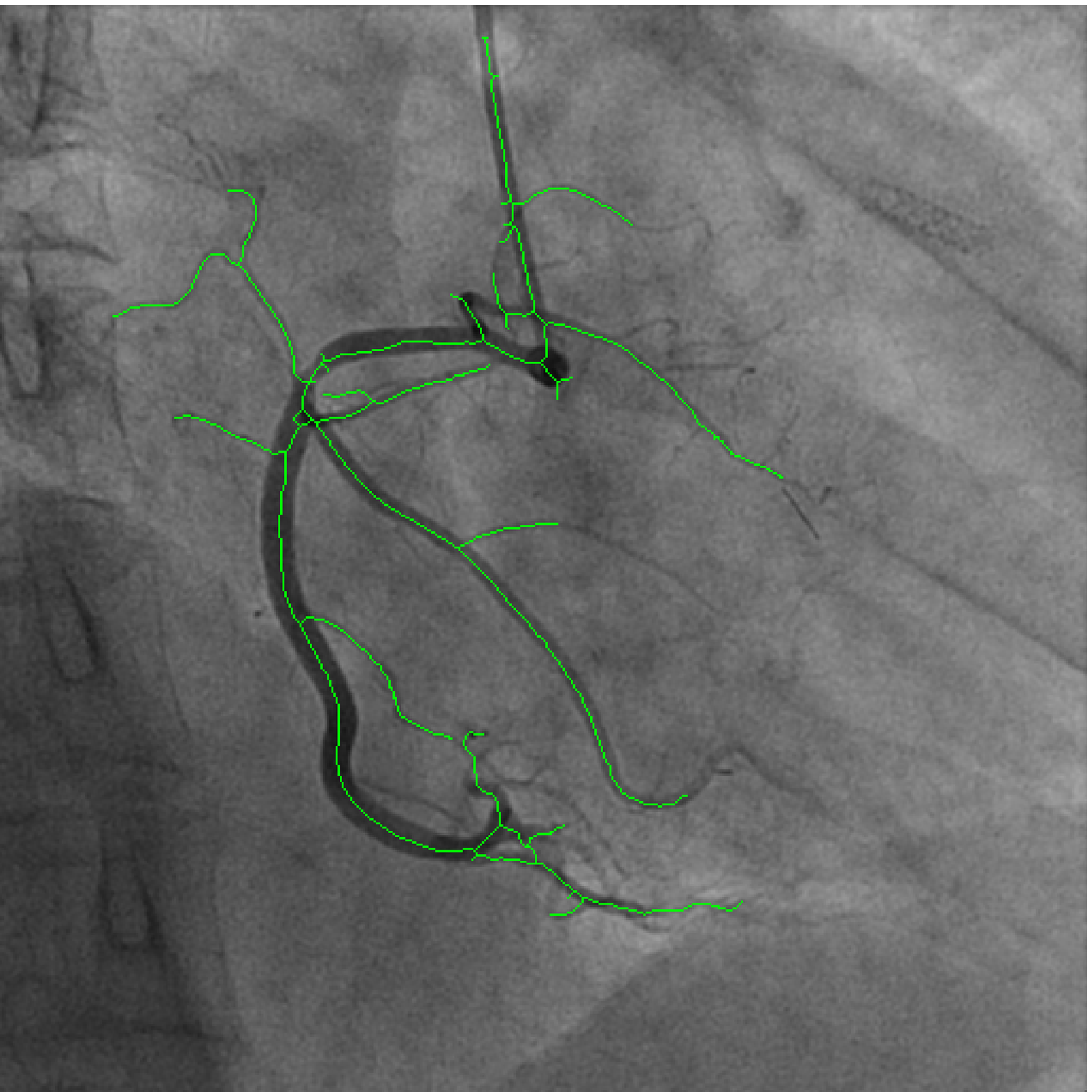}}
\subfloat{\includegraphics[width = 0.24\linewidth]{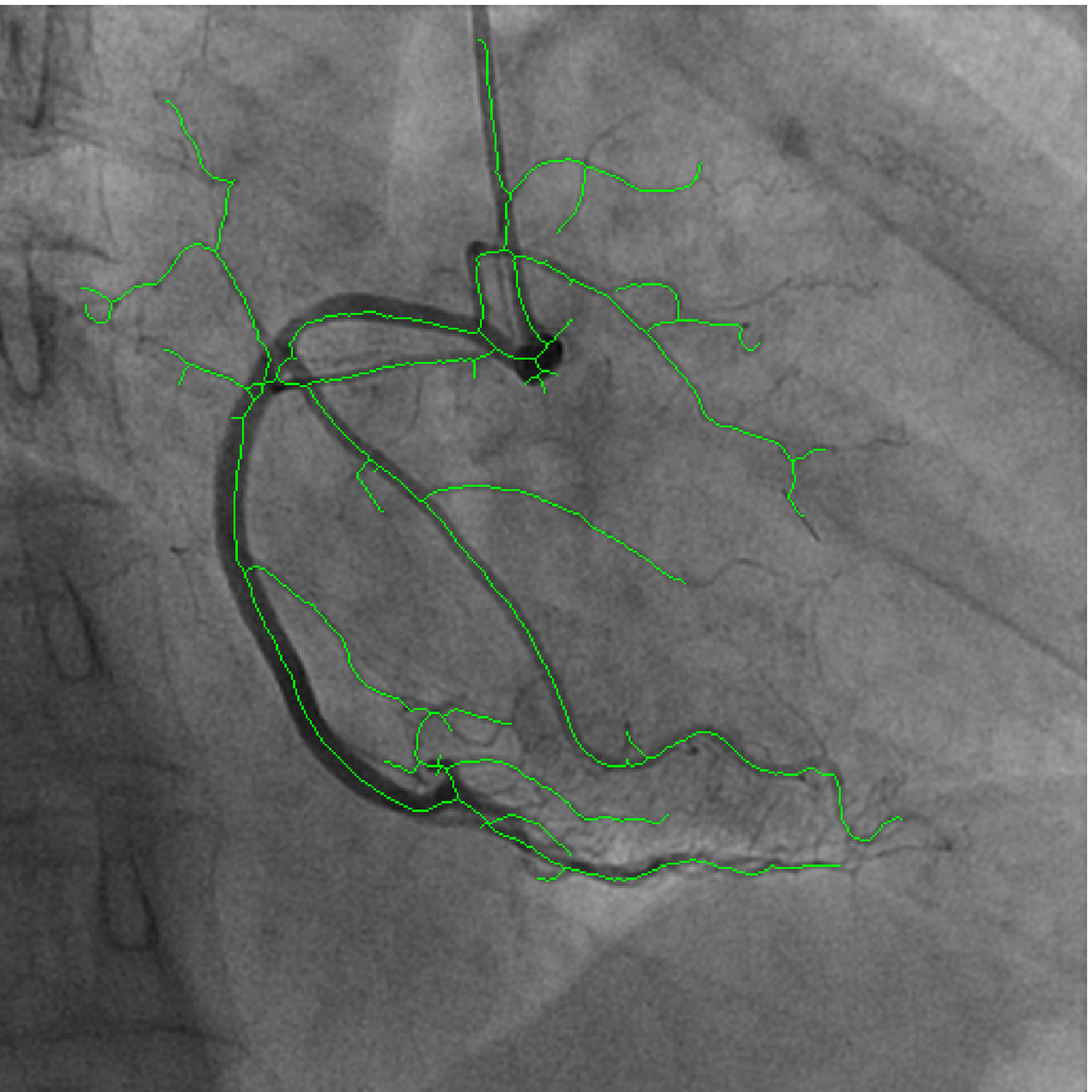}}
\caption{A fluoroscopic XRA sequence with vessels extracted by the proposed method overlaid (green). Frames (13, 18, 24, and 31, respectively) are sparsely sampled to clearly show dynamics. All figures best viewed in color.}
\label{fig:example}
\end{figure}

Fluoroscopic X-ray angiograms (XRA, Fig.~\ref{fig:example}) are used to evaluate stenosis in coronary arteries and provide guidance for percutaneous coronary intervention. Here, vessel extraction enables registration of pre-operative CT angiograms (CTA) for visualization of 3-D arterial structure. For chronic total occlusion, this can visualize otherwise invisible arteries due to blockage of contrast agent.


Many works focus on vessel extraction from a single image. Pixelwise enhancement~\cite{frangi98}, and segmentation methods with sophisticated optimization~\cite{poon07,honnorat11a} or learning~\cite{becker13} are some examples. While these methods are applicable to a wide variety of vessels, they do not consider temporal continuity and thus may give inconsistent results for a sequence. Many works use an accurate vessel structure extracted from a detailed 3D CTA to extract an accurate and consistent vessel structure for XRA sequences~\cite{rivest12,sun14}. Relatively few works have been proposed that do not require 3D CTA to detect and track curvilinear structures such as vessels~\cite{fallavollita07} or guide-wires~\cite{honnorat11b} from fluoroscopic image sequences.


\begin{figure}[t]
\centering
\subfloat[]{\includegraphics[width = 0.2\linewidth]{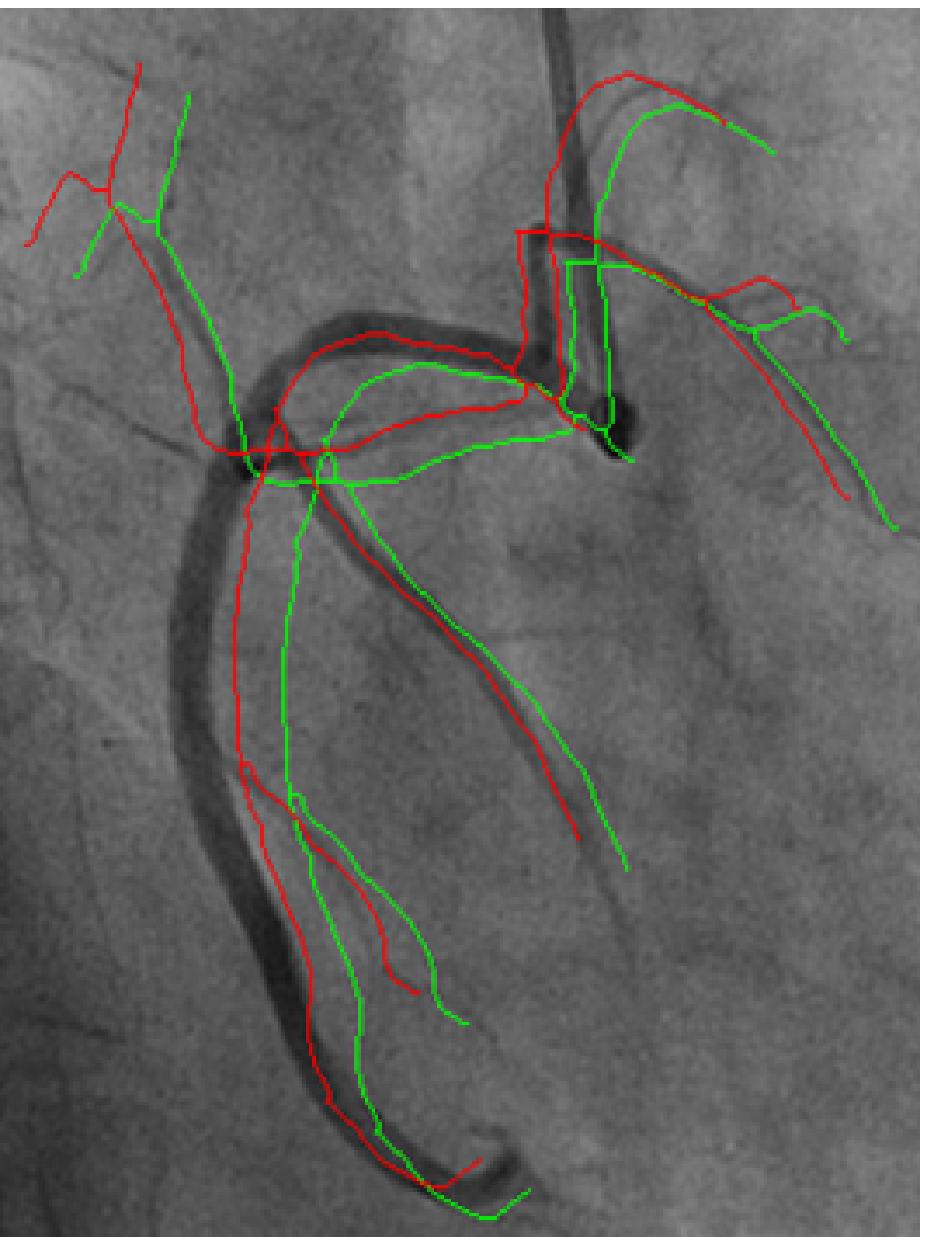}}
\subfloat[]{\includegraphics[width = 0.2\linewidth]{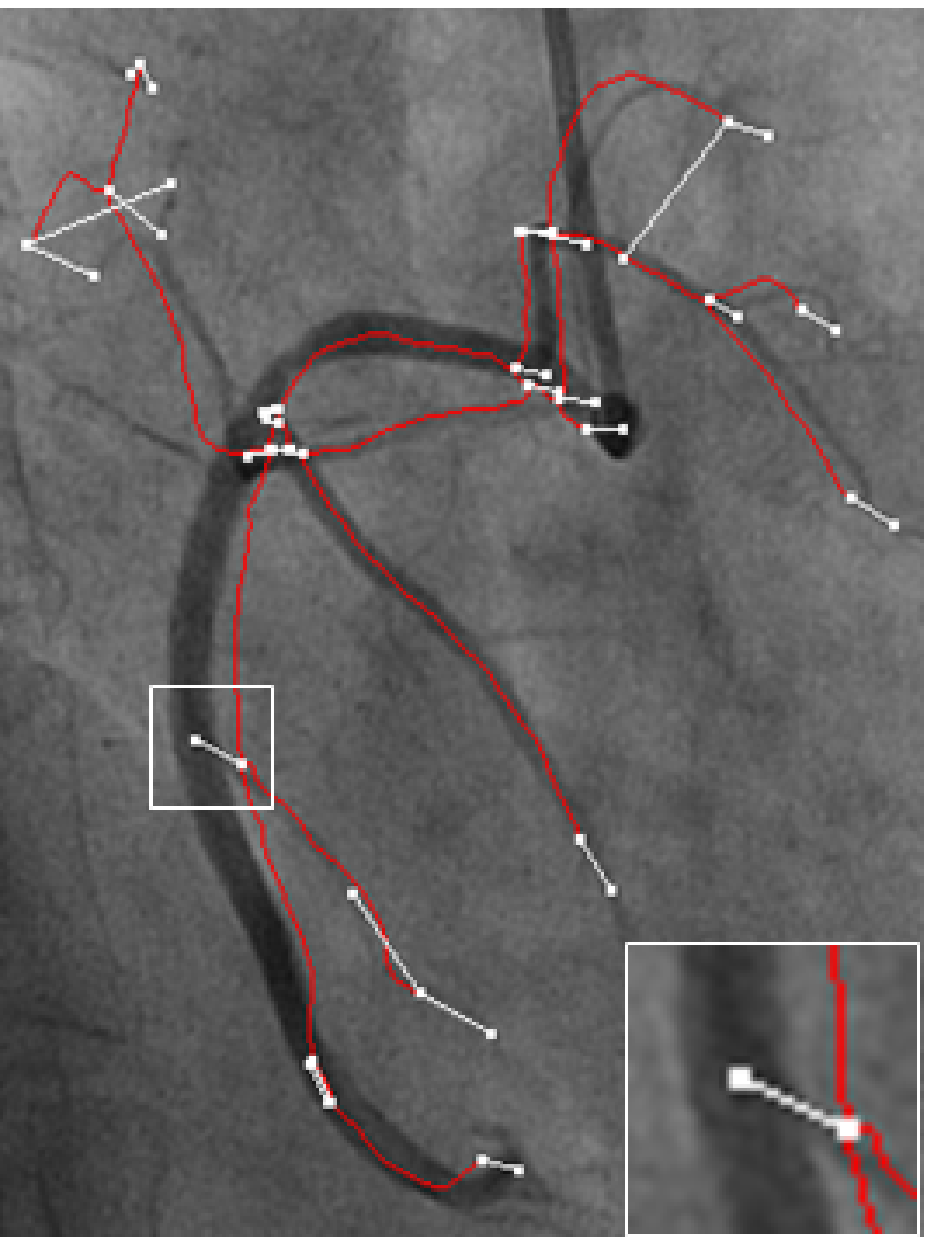}}
\subfloat[]{\includegraphics[width = 0.2\linewidth]{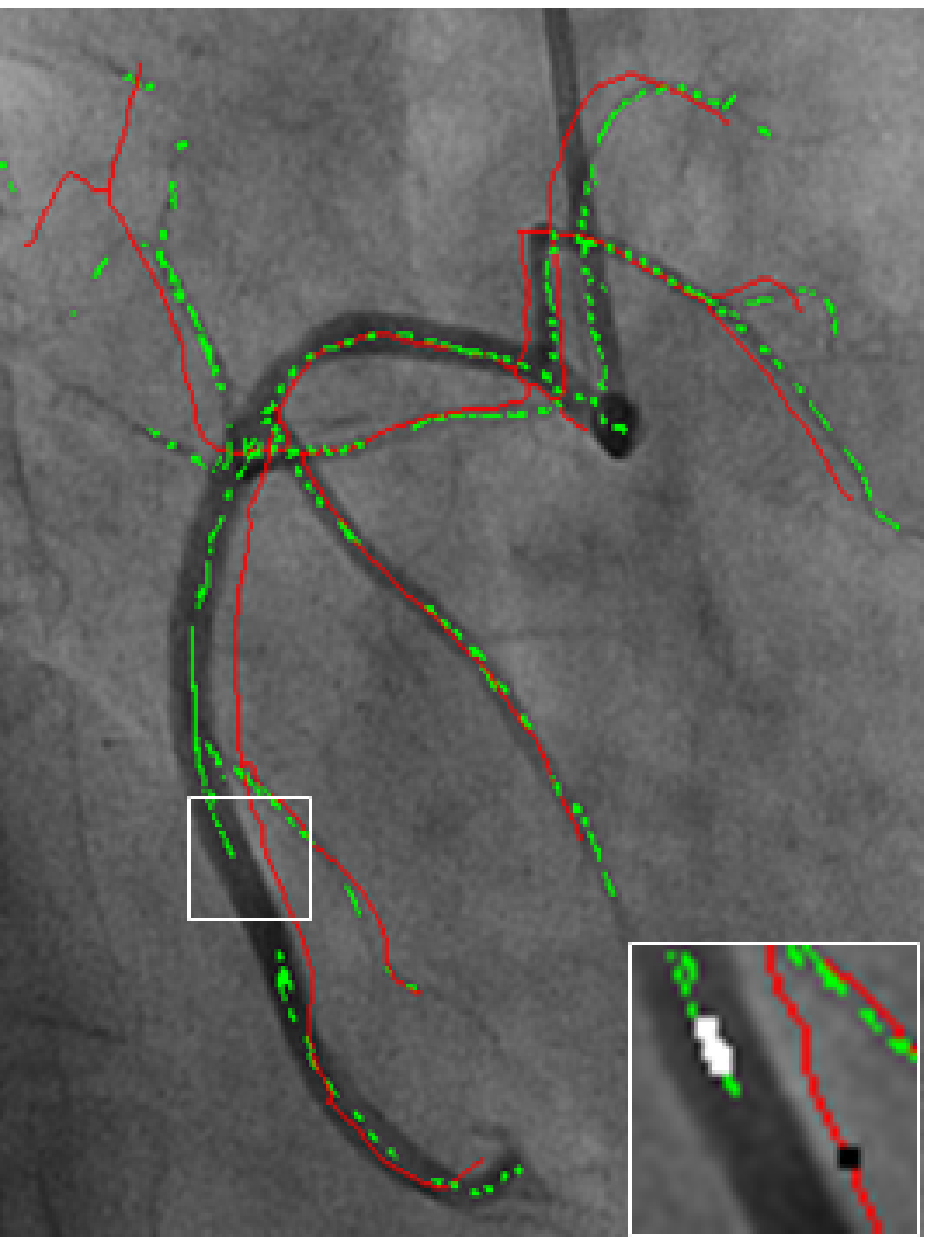}}
\subfloat[]{\includegraphics[width = 0.2\linewidth]{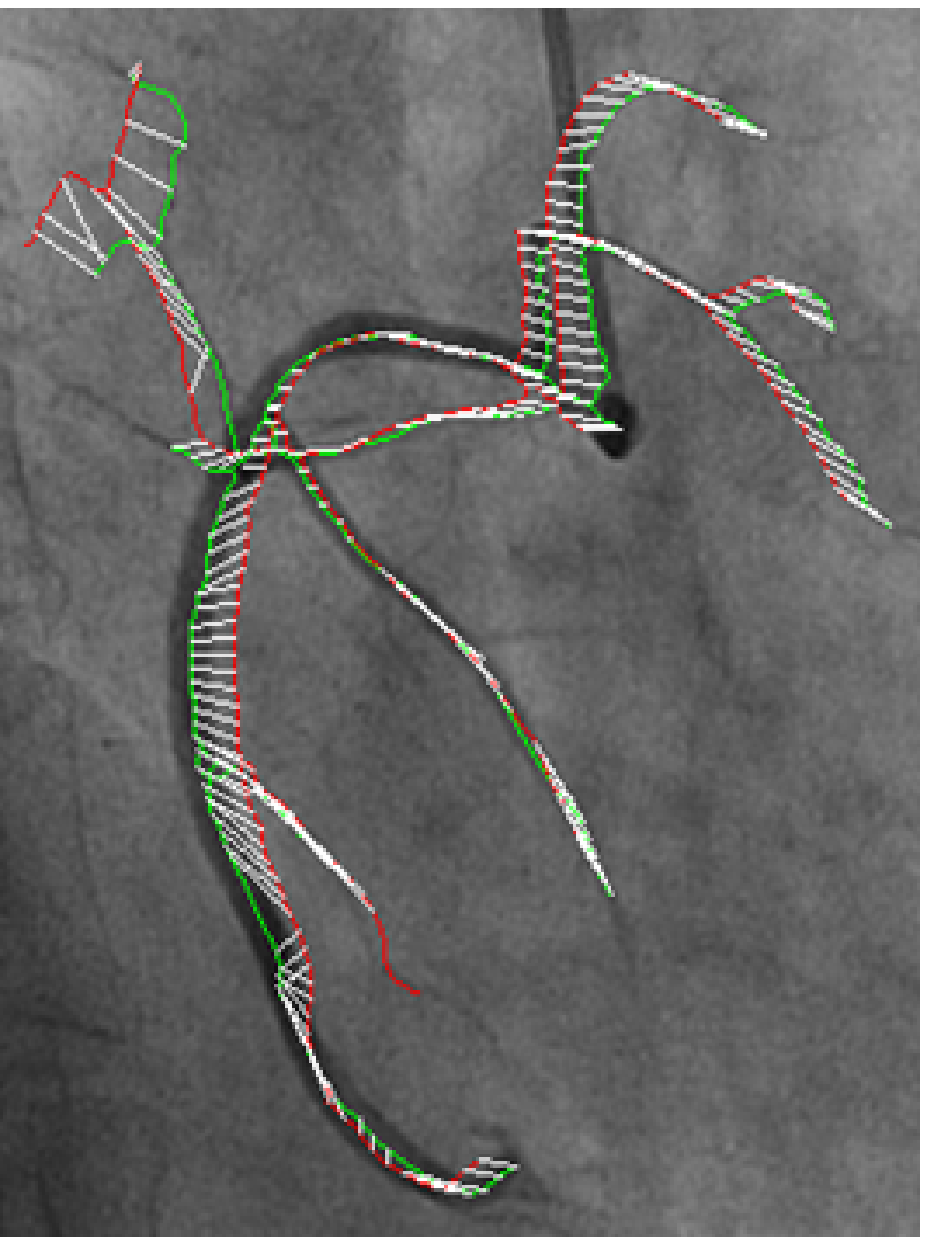}}
\subfloat[]{\includegraphics[width = 0.2\linewidth]{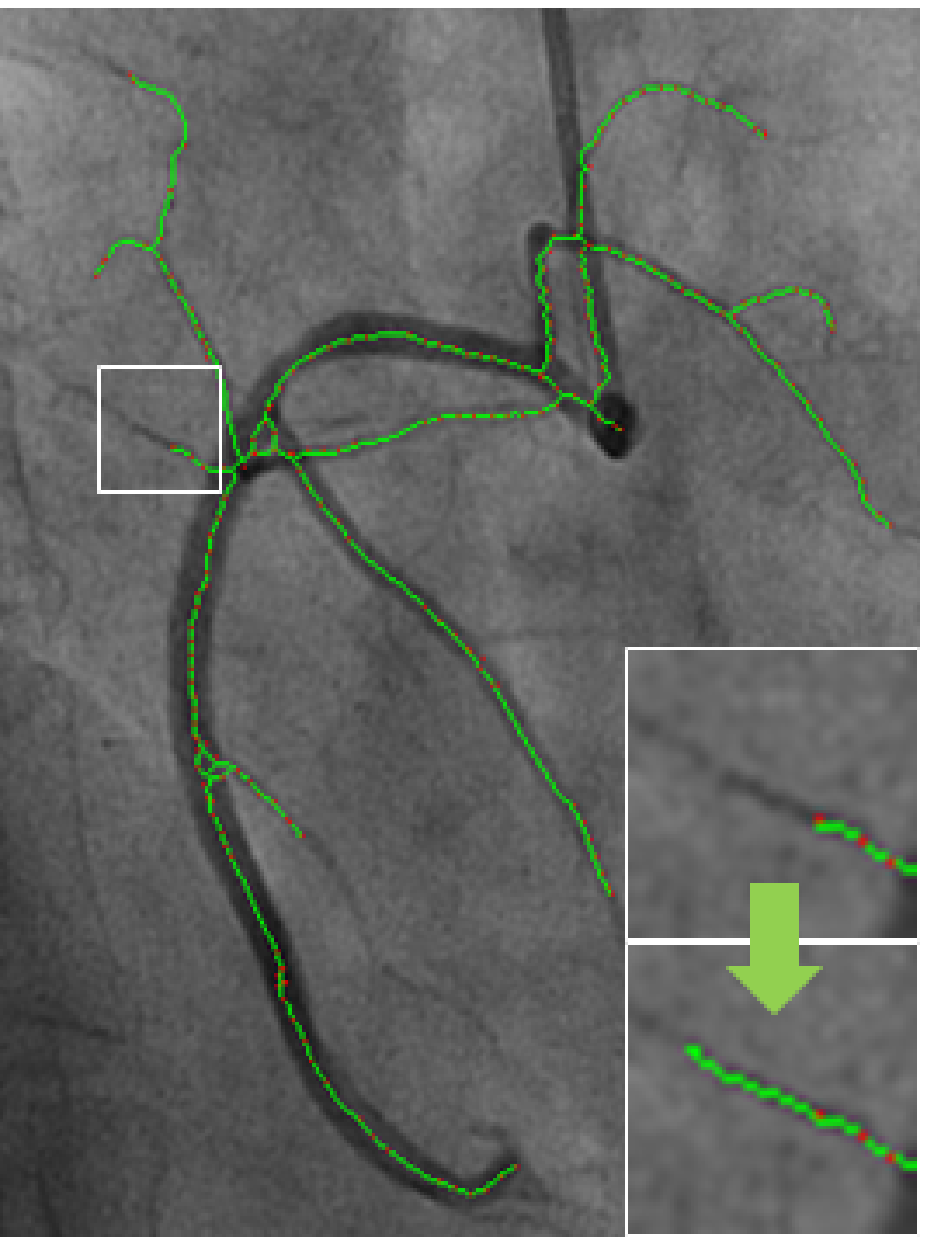}}
\caption{Illustration of overall framework. Vessel centerlines of the destination frame are extracted based on the centerlines of the source frame, which we assume to be given. (a) Global search by chamfer matching. Source (green) and translated (red) vessel centerlines are overlaid. (b) Vessel branch search by keypoint correspondence (white). (c) Correspondence candidates (green) by vessel point search. White points in zoomed box show candidates for the black vessel point. (d) Optimal point correspondences (white) from MRF optimization. (e) Extraction of newly visible vessel branches. (a)-(c) comprise a hierarchical search scheme.}
\label{fig:flow_chart}
\end{figure}

Thus, we present a method, which we term vessel correspondence optimization (VCO), to extract coronary vessels from fluoroscopic x-ray angiogram sequences. Given the vessel structure of a source frame (obtained by manual annotation or automatic methods~\cite{poon07,honnorat11a,becker13}), the detailed global and local motion is estimated by determining the optimal correspondence for vessel points in the subsequent frame. Local appearance similarity and structural consistency are enforced within a Markov random field (MRF) optimization framework. Essentially, VCO performs registration of the vessel structure. Post-processing is performed to deal with vessel branches newly visible due to the inflow of contrast agent. The proposed method is summarized in Fig.~\ref{fig:flow_chart}.

The main contributions are the development of i) an MRF optimization method for optimal registration of the vessel structure, and ii) an accurate vessel extraction method for XRA sequences based on accurate motion estimation. Experiments show that VCO is robust to complicated vessel structures and abrupt motions. We believe that VCO can be used to provide analytic visual information to the clinician without CTA acquisition. It can also be used in the automatic registration of vessels between 2D XRA sequences and 3D CTA.

\section{Vessel Correspondence Optimization}

\subsection{Markov Random Field Representation}\label{MRF_rep}


A pairwise MRF graph is constructed from the vessel centerlines of the source frame. Nodes correspond to sampled points from the centerlines and edges represent connectivities between the vessel points. The MRF energy is defined as follows:
\begin{equation}
\label{eq:OverallEnergy}
E(\textbf{x}) = \sum\limits_{i} {\varphi (x_{i} )} + \sum\limits_{(i,j) \in \varepsilon } {\psi ( x_{i} ,x_{j} )},
\end{equation}
where $\textbf{x}$ is the vector comprising the set of all random variables $x_i$ at each node with index $i$. Each $x_i$ can be labeled by $N_p +1$ different values, and the optimal $\textbf{x}$ is determined by minimizing (\ref{eq:OverallEnergy}). The $N_p +1$ labels comprise $N_p$ correspondence candidates and $1$ dummy label which will be assigned to a node when there are no candidates with consistent local shape or appearance. If the dummy label is found to be optimal, that node is excluded from the resulting VCO point set.

The unary cost function $\varphi(x_{i})$ depends on the similarity between the local appearance of the $i$th node and the $x_i$th correspondence candidate. Since we seek corresponding points with similar appearance, we define $\varphi(x_{i})$ to decrease as local appearance similarity increases as:
\begin{equation}
\label{eq:unaryEnergy}
\varphi (x_i ) = \min (\left\| {D(F_{src}, p_i )  - D(F_{dst}, \pi_i (x_i ) ) } \right\|,T_\varphi  ).
\end{equation}
$F_{src}$ and $F_{dst}$ denote the source and destination frames, $p_i$ and $\pi_i (x_i )$ denote the coordinate of the $i$th node of the source vessel structure and the $x_i$th correspondence candidate from the destination frame, respectively. $D$ is a function for a local feature descriptor. Note that $\varphi (x_i )$ is truncated by $T_\varphi$ to ensure robustness to outliers. Outliers may occur when there is no corresponding point due to severe local deformations.

The pairwise cost $\psi (x_i ,x_j )$ enforces similar displacement vectors between neighboring points, and thus consistent local shape. It is defined similar to that of \cite{glocker07}, as follows:
\begin{equation}
\label{eq:pairwiseEnergy}
\psi (x_i ,x_j ) = \lambda  \min (\left\| { ( p_i - \pi_i (x_i ) ) - ( p_j - \pi_j (x_j ) ) } \right\|, T_\psi  ),
\end{equation}
where $p_i$ and $p_j$ are coordinates of the $i$th and $j$th source node, while $\pi_i (x_i )$ and $\pi_j (x_j )$ are coordinates in the destination frame of the correspondence candidates of $x_i$ and $x_j$, respectively. $p_i - \pi_i (x_i )$ is the displacement vector of the $i$th source node. Again, truncation is included based on threshold $T_\psi$. The parameter $\lambda$ controls the amount of this regularization in (\ref{eq:OverallEnergy}).





\subsection{Hierarchical Search of Vessel Correspondence Candidates}

The tubular shapes of vessels, together with the aperture problem, make it very challenging to distinguish different local regions. We thus propose a hierarchical correspondence search scheme comprising global, branch, and point searches.

We define vessel junctions, including bifurcations and crossings from 3D-2D projection, and endpoints, both of which have distinctive appearances, as vessel keypoints. A vessel branch refers to the line connecting two vessel keypoints. In the following, we denote the $\alpha$th keypoint as $p^\alpha$, with a superscript, to distinguish it from general vessel point $p_i$. The $m$th branch is denoted as $b_m$.

\textbf{Global Search by Chamfer Matching}:
We perform chamfer matching~\cite{barrow77} to estimate large global translational motion from heart beating, breathing, or viewpoint change. The template shape is the set of source vessel points. The target shape is constructed by sequentially applying vessel enhancement~\cite{frangi98}, thresholding, and skeletonization to the destination frame. We find the global displacement vector that minimizes the sum of distances between each template point and target shape by brute force search on the distance transform (DT) of the target shape. Fig.~\ref{fig:flow_chart}(a) shows an example result of this step. 

\textbf{Branch Search by Vessel Keypoint Correspondence}:
We search for corresponding points $\pi^\alpha$ and $\pi^\beta$ for both keypoints $p^\alpha$ and $p^\beta$ of a branch $b_m$, each within a local search region of size $w_{k} \times h_{k}$. Correspondences are determined by similarity of local appearance, measured using (\ref{eq:unaryEnergy}). Non-max suppression is applied to avoid nearby matches, and up to $N_k$ possible correspondences are obtained for both $p^\alpha$ and $p^\beta$. The set of candidate branches is generated for $b_m$ by simply applying all displacement vectors $\delta^\alpha_{1} = \pi^\alpha_{1}-p^\alpha, ... ,\delta^\alpha_{N_{k}} =\pi^\alpha_{N_{k}}-p^\alpha $ and $ \delta^\beta_{1} = \pi^\beta_{1}-p^\beta, ... ,\delta^\beta_{N_{k}} =\pi^\beta_{N_{k}}-p^\beta $ to $b_m$. We note that there can be up to $N_k \times 2$ candidate branches, depending on the number of keypoint correspondences. We also include the branch with no displacement, in case all keypoint matches are unreliable, which results in at most $N_k \times 2 + 1$ branch candidates.

\begin{figure}[t]
\centering
\subfloat[]{\includegraphics[width = 0.17\linewidth]{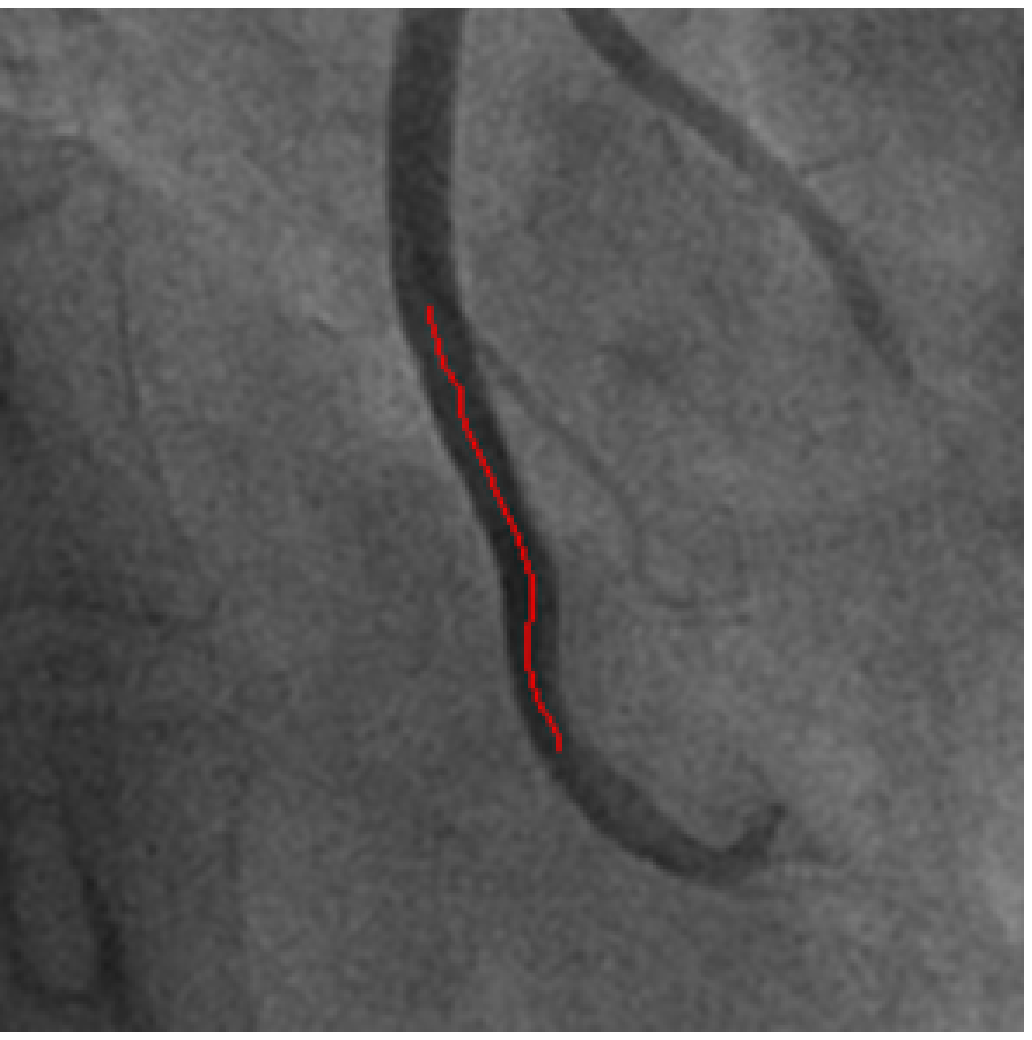}}
\hspace{0.5mm}
\subfloat[]{\includegraphics[width = 0.51\linewidth]{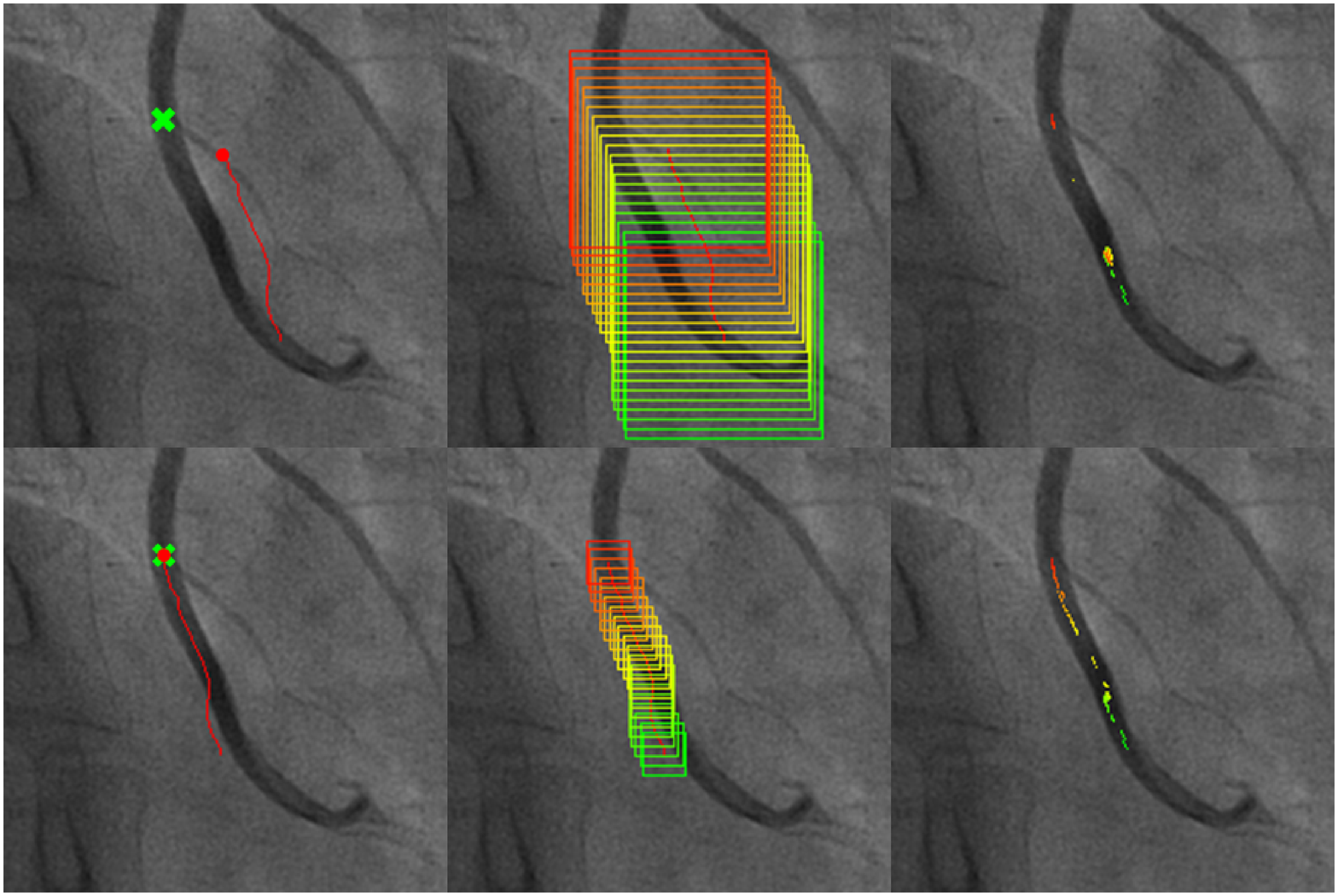}}
\caption{Effect of hierarchical branch-point search. (a) Example vessel branch in source frame. (b) Each column shows (left) the positioned vessel branch in the destination frame, (mid) local search regions at branch vessel points, and (right) obtained correspondence candidates. Top/bottom rows compare the local point search without and with hierarchical search. Branch alignment enables reduction of local search range. Search regions and candidate points color-coded for clarity. }
\label{fig:candidates}
\end{figure}

\textbf{Correspondence Candidate Generation by Vessel Point Search}:
For a vessel point $p_{i} \in b_{m}$, the set of corresponding points based on candidate branches are $\{p_{i}+\delta^\alpha_{1}, ... ,p_{i}+\delta^\alpha_{N_{k}}, p_{i}+\delta^\beta_{1}, ... ,p_{i}+\delta^\beta_{N_{k}}, p_{i} \}$. We define local search regions of size $w_{p} \times h_{p}$ at each of these points and determine the $N_l$ best corresponding points, again, based on (\ref{eq:unaryEnergy}). Fig.~\ref{fig:candidates} shows an example where the correspondence candidates are greatly improved with a smaller local search range based on the prior branch search.

The resulting maximum number of candidates is $N_p = N_l \times (N_k \times 2 + 1)$. Due to non-max suppression, the actual candidate number $N_c$ can be less than $N_p$ depending on the image, which can complicate implementation. Thus, we fix the number of labels to $N_p$ for all vessel points, but nullify labels larger than $N_c$ without an actual corresponding candidate by assigning an infinite unary cost.

\subsection{Post-processing for Complete Centerline Extraction}

After VCO, obtained vessel points are connected to construct vessel centerline structure $\mathcal V$ by the fast marching method~\cite{yatziv05}. To extract newly visible branches, a binary segmentation mask is first obtained by thresholding vesselness~\cite{frangi98}, and regions not connected to the vessel centerlines are excluded. The DT is computed for this mask, with the non-vessel regions as seeds. Based on the mask and the DT, the following is iterated until no branch longer than the maximum vessel radius is found: i) perform fast marching method, with $\mathcal V$ as the seeds and the DT values as marching speed; ii) find the shortest path from the pixel with the latest arrival time to $\mathcal V$, and add to $\mathcal V$. This method is adopted from \cite{van07}.


\section{Experimental Results}

\textbf{Evaluation Details }:
The dataset comprises 18 XRA sequences of total 617 frames from 5 different patients. All sequences were acquired by Philips digital fluoroscopic systems at $512 \times 512$ resolution, 8 bit depth, and 15 fps frame rate. Parameter values were manually selected and fixed to $N_k=2$, $w_{k}=101$, $h_{k}=101$, $N_l=5$, $w_{p}=21$, $h_{p}=21$, and $N_{p} = N_l \times (N_k \times 2 + 1) = 25$. Source vessel points are sampled from the centerline at a 5 pixel interval. The VLFeat~\cite{VLFeat} library for SIFT~\cite{lowe04} is used for $D$ in (\ref{eq:unaryEnergy}). TRW-S is used for MRF optimization~\cite{kolmogorov06}. VCO without post-processing took a few minutes by an unoptimized Matlab implementation on a Intel Zeon processor with over $80\%$ of computation on SIFT matching. Ground truth vessel centerlines and corresponding bifurcation point coordinates in frame pairs were obtained by expert annotation. For centerlines, the semi-automatic method of \cite{poon07} was used.


Comparisons are made with two relevant methods: 1) that by Fallavollita et al.~\cite{fallavollita07}, which combines optical flow (OF) and an active contour model (ACM), modified to handle whole vessel structures, denoted as \textbf{OF+ACM}, and 2) a variant of \textbf{OF+ACM} where OF is substituted with global chamfer matching, denoted as \textbf{CM+ACM}. Further comparisons are made with two variants of VCO: 3) \textbf{VCO-HS} VCO without hierarchical search, with enlarged search regions instead, and 4)\textbf{VCO-DL}, VCO without dummy labels.

\textbf{Quantitative Evaluation}:
We perform two different experiments. In \textbf{Exp1}, vessel extraction is performed for all source-destination frame pairs, using the ground truth source vessel structure. The average precision, recall, F-measure, along with the target registration error (TRE) for bifurcation points, of 599 sequential frame pairs are presented in Table.~\ref{quan_res}. TRE is defined as the average distance between the estimated and ground truth point coordinate. A vessel point is true positive if there is a ground truth point within a two pixel radius. VCO achieves the highest accuracy, with additional improvement from hierarchical search and dummy labels. Here, post-processing is applied only up to point connection to exclusively evaluate the VCO accuracy.

In \textbf{Exp2}, we evaluate the average number of consecutive frames with \emph{sufficient} vessel extraction when iteratively applying all methods for a single initial frame. Here, \emph{sufficient} is defined as F-measure higher than $0.7$. Evaluation on the 18 sequences showed that \emph{sufficient} vessel extraction was obtained for 2.0, 5.3, and 8.6 subsequent frames, by the \textbf{OF+ACM}, \textbf{CM+ACM}, and \textbf{VCO} methods, respectively. This demonstrates the practical usefulness of VCO.


\begin{table}[t]
\begin{adjustwidth}{}{}
\centering
\caption{Quantitative results of \textbf{Exp1}, where vessel extraction is performed for 599 frame pairs using the ground truth source vessel structure. Higher is better for precision, recall, and F-measure, and lower is better for TRE.}
\label{quan_res}
\begin{tabular}{|c|c|c|c|c|c|}
\hline
          & OF+ACM~\cite{fallavollita07} & CM+ACM & VCO-HS & VCO-DL & VCO   \\ \hline
Precision & 0.832               & 0.860         & 0.900                 & 0.901               & 0.905 \\ \hline
Recall    & 0.733               & 0.719         & 0.840                 & 0.841               & 0.841 \\ \hline
F-measure & 0.779               & 0.783         & 0.869                 & 0.870               & 0.872 \\ \hline
TRE       & 6.321               & 6.418         & -                   & -                 & 5.018 \\ \hline
\end{tabular}
\end{adjustwidth}{}{}
\end{table}

\begin{figure}
\centering
\includegraphics[width = 0.83\linewidth]{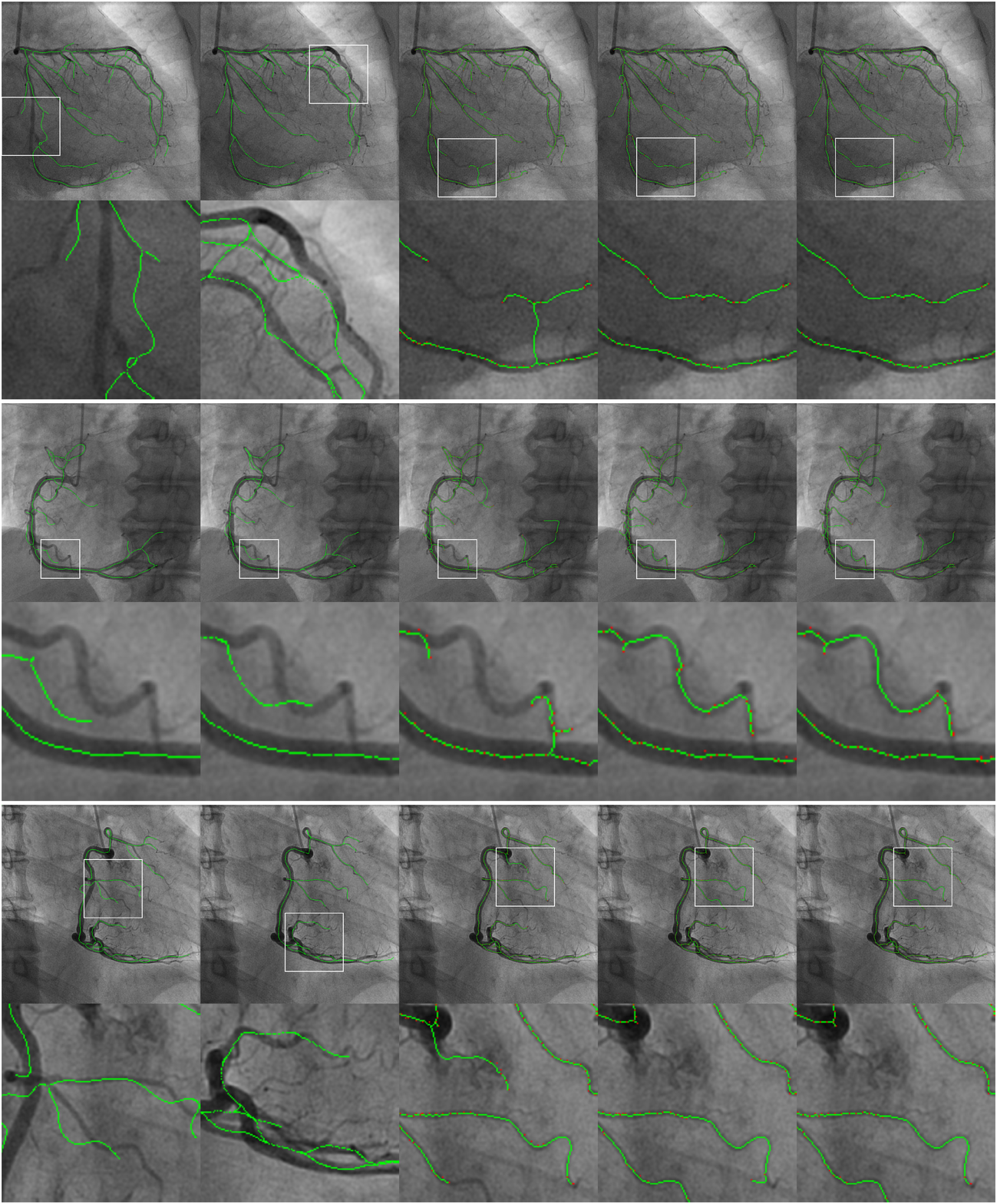}\\
\begin{minipage}[b]{.166\linewidth}
\centering{OF+ACM~\cite{fallavollita07}}
\end{minipage}%
\begin{minipage}[b]{.166\linewidth}
\centering{CM+ACM}
\end{minipage}%
\begin{minipage}[b]{.166\linewidth}
\centering{VCO-HS}
\end{minipage}%
\begin{minipage}[b]{.166\linewidth}
\centering{VCO-DL}
\end{minipage}%
\begin{minipage}[b]{.166\linewidth}
\centering{VCO}
\end{minipage}%
\caption{Qualitative results. Odd and even rows show sample frames and their corresponding enlarged views of erroneous regions, from different sequences, respecitively. Red points in column 3-5 are resulting vessel points of VCO.}
\label{fig:qual_res}
\end{figure}
\begin{figure}
\centering
\subfloat[Vessel superimposition]{\includegraphics[width = 0.42\linewidth]{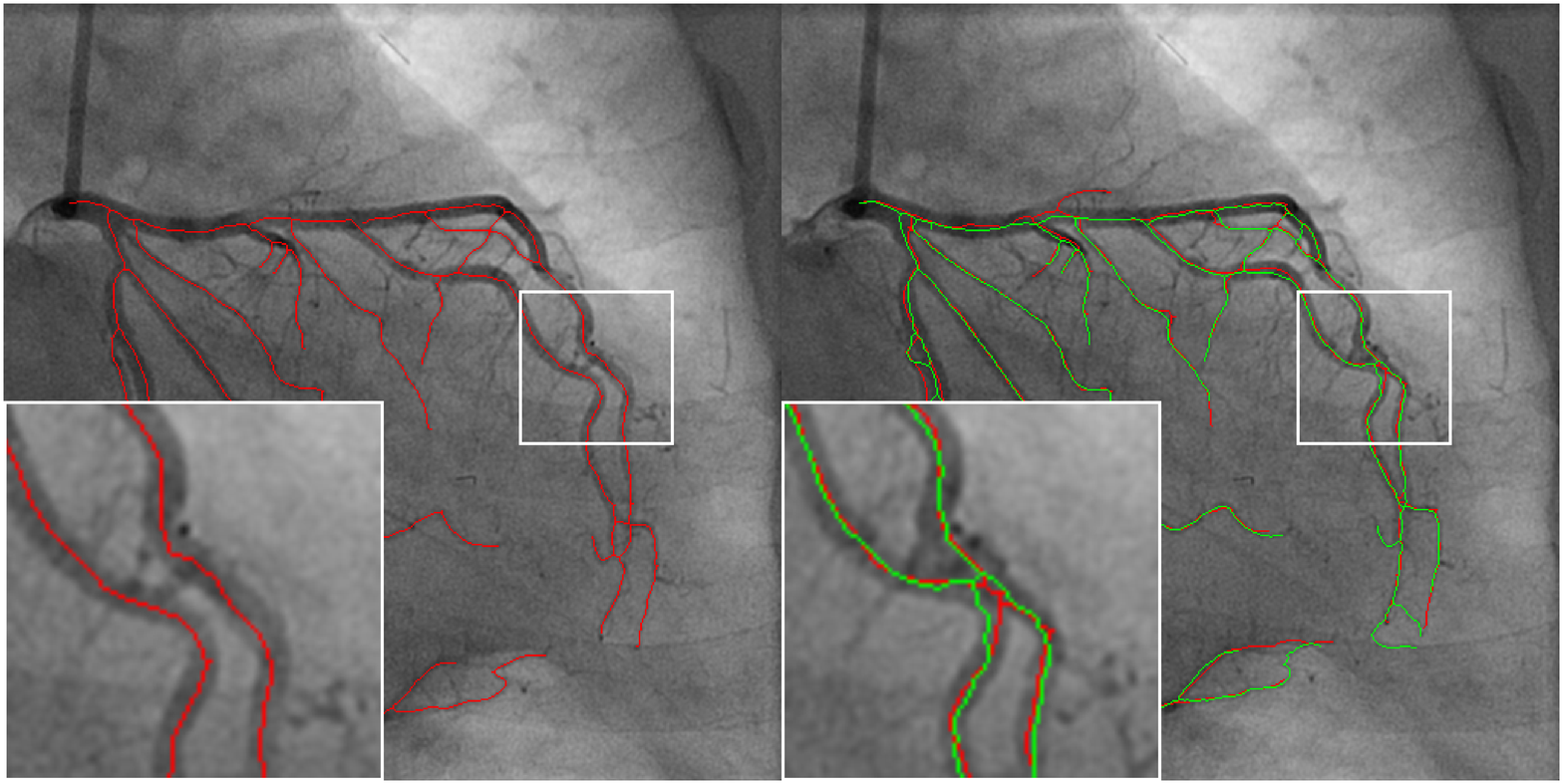}}
\hspace{0.5mm}
\subfloat[Vessel separation]{\includegraphics[width = 0.42\linewidth]{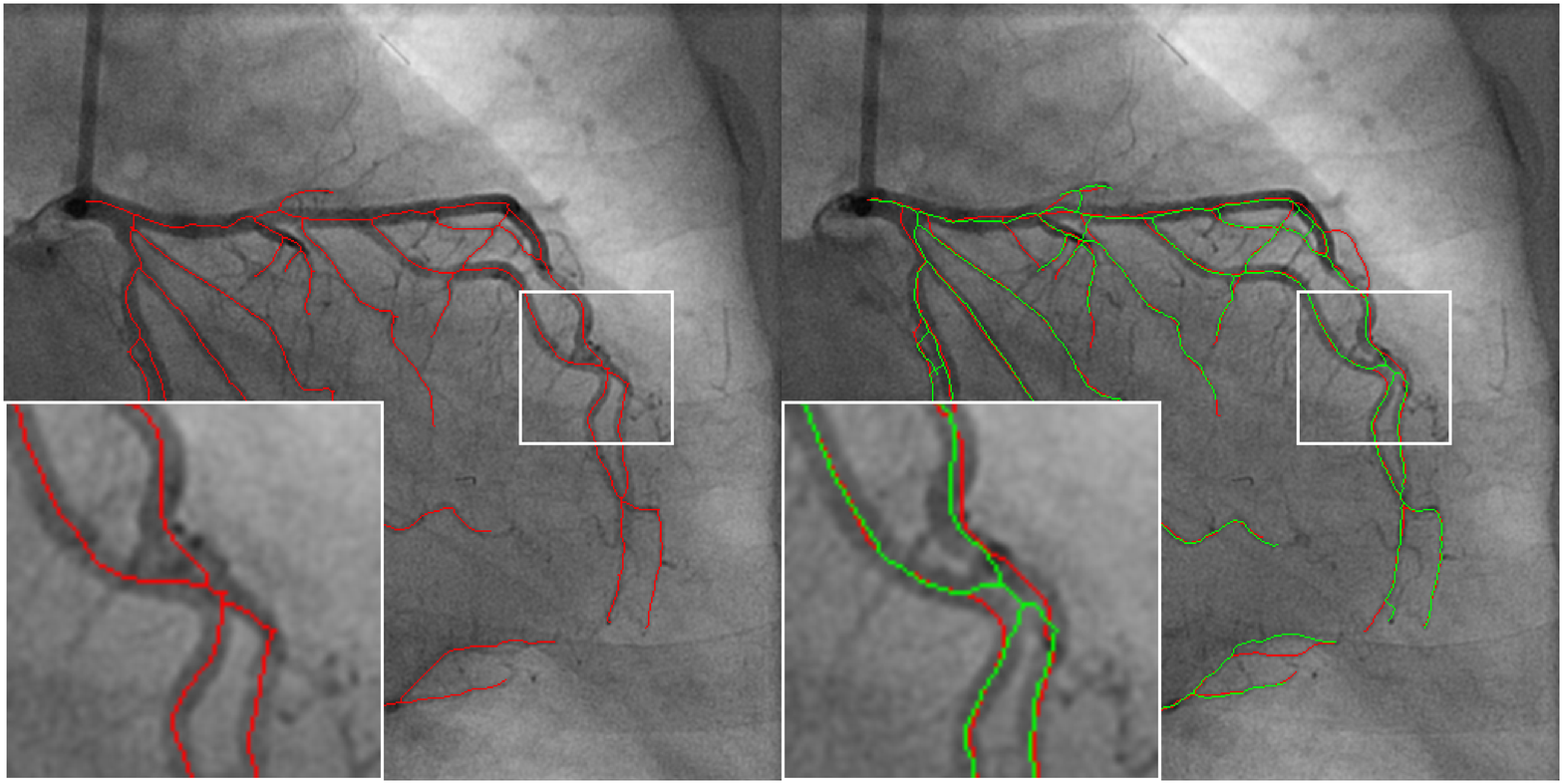}}
\caption{Limitations of VCO. Subsequent frame pairs showing (a) success and (b) failure. Left/right show source/destination frame pairs. Ground truth (red) and VCO result vessel structures (green) are overlaid.}
\label{fig:super_case}
\end{figure}

\textbf{Qualitative Evaluation}:
Fig.~\ref{fig:qual_res} presents representative sample results of \textbf{Exp1}. Significant improvements compared to previous methods are visible. Fig.~\ref{fig:example} presents results of \textbf{Exp2} for a sample sequence. Both figures highlight the effectiveness of the proposed method. Fig.~\ref{fig:super_case} shows one limitation of VCO, where VCO is not able to handle the topology change due to vessel superimposition. 

\section{Conclusion}

We have proposed a method to determine optimal vessel correspondences for registration and extraction of vessel structures from fluoroscopic x-ray sequences. Experiments show promising results on complicated structures. In future work, we plan to investigate optimization measures including GPU implementation for real-time performance as well as other measures required for actual clinical application such as overlaying the extracted vessel structure to sequences acquired without the use of contrast agents.


%
%
%
%
%

\end{document}